\documentstyle[12pt]{article}
\def\title#1{\begin{center}{\Large{\bf #1}}
\end{center}}
\def\authoraddress#1#2{\begin{center} #1 \\
{\it #2 }\end{center}}
\def\keywords#1{\begin{flushleft}
{\large KEYWORDS:}
#1\end{flushleft}}

\def\preprint#1{
\hfill
\begin{flushright}
#1
\end{flushright}
}
\def\eline{\vfill
\begin{picture}(300,5)
\put(0,0){\line(1,0){280}}\end{picture}\\}

\def\acknowledgement#1{
\vspace*{1cm}
\leftline{\bf Acknowledgement}\par
#1
}

\def\eqn#1{\begin{eqnarray}#1\end{eqnarray}}
\def\non{\nonumber}

\def\disp{\displaystyle}
		\def\De{\Delta}

\begin{document}
\begin{titlepage}
\preprint{OUCMT-94-3}
\title{
Dynamical Phase Transition in\\
One Dimensional Traffic Flow Model with Blockage${}^\dagger$
}
\authoraddress{
Satoshi YUKAWA${}^{\ddag}$, Macoto
KIKUCHI${}^{\ddag \ddag}$ and Shin-ichi TADAKI${}^{1 \ast}$
}{
Department of Physics, Faculty of Science, Osaka university \\
Toyonaka 560 \\
${}^1$Department of Information Science,Saga University \\
Saga 840\\
}


\begin{abstract}

Effects of a bottleneck in a linear trafficway is investigated
using a simple cellular automaton model.
Introducing a blockage site which transmit cars at some
transmission probability into the rule-184 cellular automaton,
we observe three different phases with increasing car concentration:
Besides the free phase and the jam phase,
which exist already in the pure rule-184 model,
the mixed phase of these two appears at intermediate concentration
with well-defined phase boundaries.
This mixed phase, where cars pile up behind the blockage to form a
jam region, is characterized by a constant flow.
In the thermodynamic limit, we obtain the exact expressions for
several characteristic quantities
in terms of the car density and the transmission rate.
These quantities depend strongly on the system size
at the phase boundaries;
We analyse these finite size effects based on the finite-size
scaling.

\end{abstract}
\eline
{\small $\dagger$ to appear in J. Phys. Soc. Jpn.} \\
\ddag {\small e-mail:yuk@bopper.phys.sci.osaka-u.ac.jp }\\
\ddag \ddag {\small e-mail:kikuchi@godzilla.phys.sci.osaka-u.ac.jp } \\
$\ast$ {\small e-mail:tadaki@ai.is.saga-u.ac.jp }
\end{titlepage}
\section {Introduction}\label{sec:intro}
\setcounter{equation}{0}

Traffic flow problems have become common social problems.
Theoretical study of traffic flow has been mainly
based on the methods of fluid dynamics.\cite{L88}
In their pioneering work, Lighthill and Whitman have given
theoretical arguments on the traffic flow in linear roads
using fluid-dynamical ideas;\cite{LW55}
When the car concentration is very low,
each car can move with the maximum speed allowed
so that the average speed is constant irrespective of the
concentration.
However, if the concentration exceeds some limiting value,
some of the cars have to lower their speed;
thus the average speed decreases with increasing concentration.
This limiting concentration may be regarded as the onset of the
traffic jam.
One of the important concepts for discussing the traffic jam
is the capacity of the road.
In ref.2, the capacity is defined through the
relation between the concentration and the average flow
which is given by the average speed times the concentration.
For relatively low concentration,
the average flow increases with the concentration.
On the other hand, in very crowded situation,
the average flow decreases with the concentration.
Therefore at some value of the concentration, the flow
takes its maximum;
that value of the concentration gives the capacity of the road.

Occurrence of the traffic jam is not the only reason
why the traffic flow has been attracted attention in physics.
For example, the $1/f$ fluctuation observed in the traffic flow
has also been studied both experimentally and theoretically.
\cite{HM,TT93}

Quite recently, cellular automata (CA) have become used for modeling
traffic flow.${}^{5-9)}$
One of the simplest models describing the traffic flow on a linear
road is the rule-184 elementally CA according to the naming scheme by
Wolfram.
\cite{wolf}
In contrast to the fluid dynamical modeling,
cars are treated as distinguishable particles in CA models.
In general CA models, the roads are expressed by discrete lattices,
and the systems evolve in discrete time step;
Owing to this space-time discritization,
the CA models are easy to be tracted by computer simulations.
In case of the rule-184 CA,
each particle can move to the next site in one direction at each time
step; All the particles whose destination sites are empty move
simultaneously.
Thus, the time evolution of this CA is completely deterministic.
Although it is a very simple rule,
it exhibits a sharp phase transition between the freely-moving phase
and the jam phase;
The onset of the jam and the capacity coincide with each other
in this model.
It is known, however, that this phase transition is trivial
in a sense that no particular phenomena, such as critical phenomena,
is observed at the transition point.
Yet it is useful as a starting point for studying the properties of
the traffic flow.
The rule-184 cellular automaton may also be regarded as
a simplified discrete version of the Burgers equation.\cite{burgers}

In this paper, we study effects of a bottleneck on the traffic flow
in a linear road.
The bottleneck is defined as a part of the road
whose capacity is lower than that of the rest of the road.
As has already been discussed in ref.~2,
it is expected that if the concentration is high to some extent
the cars pile up behind the bottleneck and form a ``shock front''.
We introduce a blockage site in the rule-184 CA for
take the bottleneck into account.
The blockage site transmit the cars with some transmission
probalility. This blockage mimics the effect of the bottleneck
caused by the road construction, the tunnel, and so on
in the real world. As will be shown in the later sections,
the introduction of the blockage site causes nontrivial behavior.
Quite recently, Nagatani has investigated the effects of a bottleneck
in two-lane trafficway by a CA model in which cars can move back and
forth stochastically between the two lanes.\cite{N94}
The model we treat, on the other hand, has only a single lane,
and the movements of the cars are deterministic except a single site,
that is, blockage.

The CA model of traffic flow is closely related to
the one-dimensional totally asymmetric simple-exclusion
process(TASEP).\cite{L75}
The TASEP is not a deterministic process;
Rather, a randomly chosen particle is moved to its neighboring site
at each time step.
While the rule-184 is regarded as a model for the traffic flow,
the TASEP may be regarded as a model for charged particles adsorbed
on a solid surface with the electric field gradient imposed.
Janowsky and Lebowitz have reported the simulation
of TASEP with a blockage similar to one in our mode.\cite{JL92}
They found a formation of the shock front, and analysed its
properties.
Our model can also be regarded as a deterministic version of their
model. Apart from the difference in the dynamics of two models,
we focus mainly our attention on the behavior near the transition
point between the free phase and the phase where the shock front exists.

The present paper is organized as follows:
In \S~\ref{sec:dynamics}, we describe the model.
and define quantities we observe.
The results of simulations are given in
\S~\ref{sec:results}.
We perform a mean-field-like approach for thermodynamic limit,
and also make analysis based on the finite-size scaling idea.
The last section is devoted to summary and discussions.

\section{Model Dynamics}\label{sec:dynamics}
\setcounter{equation}{0}

We start from describing the pure rule-184 CA without a blockage.
Suppose we have a linear lattice where each site can take either of
the two states, 0 and 1. The CA evolves in discrete time step;
The state of each site at the next time step is determined from
the state of the site itself and those of the two nearest-neighbor
sites. The evolution rule for the rule-184 CA is expressed symbolically
by the following set of fractions:
\eqn{\left\{
 \frac{111}{1}, \frac{110}{0}, \frac{101}{1}, \frac{100}{1},
\frac{011}{1},
\frac{010}{0}, \frac{001}{0}, \frac{000}{0} \right\} \label{eq:rule}.
}
The three binary numbers in the numerators
express the states of the three sites,
the site in concern and the two neighboring sites.
These fractions express that a state, 1 or 0 in denominators,
of each site after one time step is determined
by the states of the three sites.
Now We consider the state 1 is an empty site or hole,
and the state 0 a site where a particle exist,
Then the particle is driven left by the evolution rule,
whenever the left nearest-neighbour site is empty.
The number of the particles is conserved throughout the evolution
process.
The combination of the binary number in the denominators, 10111000
is 184 in decimal number;
That is why this CA is called the rule-184 after Wolfram.\cite{wolf}
It should be noted that the rule-184 is one of the ``illegal'' rules
according to Wolfram, since it lacks the spatial reflection symmetry.
This illegality, however, is the source of the asymmetric motion.

The effects of a bottleneck is taken into account
by introducing a blockage site.
This site has a transmission probability $r$;
If a particle exists on this site and its left nearest-neighbour site
is empty, then a particle moves left with probability $r$
and does not move with probability $1-r$.
On a blockage site, the evolution rule is modified as
\eqn{\left\{
\disp \frac{1101}{01},  \disp \frac{1100}{01},
\disp \frac{0101}{01},  \disp \frac{0100}{01}\right\}  &
\disp \qquad \mbox{with probability r,} \non \\
\left\{\disp \frac{1101}{10},  \disp \frac{1100}{10},
\disp \frac{0101}{10},  \disp \frac{0100}{10}\right\}  &
\qquad \mbox{with probability 1-r.}
 \label{eq:modrule}
}
In each fraction, the blockage site is the third digit from left in
the numerator,
and the denominator expresses the state of the blockage site
at the next step.
The evolution rule for other states and other sites are the same one
as the pure rule-184.
The blockage site will act as a seed of the ``traffic jam'' for the
particles.
The model thus constructed evolves mainly with the deterministic rule
(\ref{eq:rule});
only at the blockage site the stochastical rule (\ref{eq:modrule})
applys.

We study this model by computer simulations of finite lattices
with the periodic boundary condition imposed.
The present model is expected to have a steady state,
because of the periodic boundary condition
and the global conservation of the particle.
The particle on the blockage site moves with the transmission
probability $r$,
so that the ``life time'' for a particle on the blockage site
is $ 1/r$.
If a particle comes to the blockage site,
this particle will leave the blockage site after $1+1/r$ time steps
on average.
Therefore the particles is expected to pile up after the blockage
if the particle concentration exceeds $r/(1+r) $;
Thus a traffic jam region of finite thickness is expected to form.

Let us define some quantities we observe by the simulations.
These quantities are calculated after the system reaches the steady
state.

First we define the average speed $v$ as
\eqn{
v=\frac{1}{T} \sum_i \frac{n_i}{n}, \label{eq:speed}
}
where $T$ is the total time steps,
$ n_i $ is the number of particles which move at $i$-th time step,
and $n$ is the total number of particles.
Using this average speed, the flow is calculated as
\eqn{
v \frac{n}{L}=v \rho, \label{eq:flow}
}
where $L$ is the lattice length and $\rho$ is the particle
concentration.
Next we define the width of the jam created after the blockage site.
The tail of the jam is unambiguously determined
as the last site from the blockage where the particle on it is
blocked by the particle at the left nearest-neighbor site;
In other word, all the particles after the tail can move freely.
It is one of the advantage of the present deterministic model over
the stochastic TASEP,
in which the tail position (the shock front in the terminology of
TASEP)
can be determined only statistically by means of the second-class
particle.
\cite{JL92}
If the blockage is at $I$-th site and
a tail of the jam is at $J_i$-th site  at i-th time step,
the width of jam phase $h$ is defined as
\eqn{
h=\frac{1}{T} \sum_i \left( J_i-I \right). \label{eq:width}
}
The fluctuations of width $ \De h^2 $ of the jam is calculated as
\eqn{
\De h^2 =\frac{1}{T} \sum_i \left( J_i-I \right)^2 -h^2.
\label{eq:fluct}
}

\section{Results}\label{sec:results}
\leftline{3.1. {\large \sl Simulation} }
\medskip
\setcounter{equation}{0}

We made computer simulations of the model varying the system size
$L$,
the transmission rate $r$, and the particle concentration $\rho$.
In Figs.~1(a) and (b), typical results for the
average speed and the flow are plotted, respectively,
against the concentration for three transmission rates,
$r=0.3, 0.5$, and 1, where the last one corresponds to the pure
rule-184.
We took $L=99$ in this figure.
Three phases are recognized for $r=0.3$ and $0.5$.
On the other hand, only two phases exist for $r=1$,
which we call the free phase (for $\rho <1/2$)
and the fully jam phase (for $\rho >1/2$).
We call intermediate phase appears for $r=0.3$ and $0.5$
as the flow-constant phase, because of the evident reason.
The plot of the flow is symmetric about $ \rho =1/2 $ as expected
from the particle-hole symmetry of the model;
Thus the two critical points for the blockage model are also at
the symmetric positions about $ \rho =1/2 $.
{}From the typical flow in Fig.~1(b),
we see that the values of the flow in the flow-constant phase
coincide with the values of the lower critical concentration
$\rho_c$,
and that the flow is proportional to the concentration in the free
phase andthe fully jam phase.
The behaviors of the average speed and the flow are
quite similar to those find in the two-lane model.\cite{N94}

Figures~2(a) and (b) show
the typical plots for the width of the jam region and
the fluctuations of the width, respectively,
for the three transmission rates.
We find that the width of the jam region increases linearly with
the concentration in the flow constant phase.
In Fig.~2(b),
the fluctuation of width has a peak
at the concentration slightly higher than the critical point.

To see the nature of the flow-constant phase, we show typical snap
shots. Figures 3(a) and (b) show snap shots
for $L=100$ and $r=0.5$ at $\rho=0.5$ and $0.33$, respectively.
It is clearly seen that the flow-constant phase is
a mixture of the free phase before the blockage
and the fully jam phase after the blockage, as expected.
This mixed phase reminds us of the mixture of the gas phase and
the liquid phase in wetting phenomena.
The tail of the jam, that is, the interface between the free region
and the
jam region, shows a "saw tooth" pattern,
which means that the tail moves backward gradually
and abruptly jumps forward;
This behavior coincide with our experience in a real traffic jam
caused by a road construction.
Figure 3(b) is a snap shot near the critical concentration.
In this concentration,
jam clusters like droplet appears and disappears from time to time.

The transition to the flow-constant phase
is not sharp in the figures we have seen so far,
in contrast to the case of the pure rule-184 model,
where the sharp transition between the free phase and
the fully jam phase is seen even for the finite systems.
Origin of such rounding of the phase transition is attributed
to the finite-size effects.
To study the finite-size effects,
we plot $v$, $h/L$, and $\De h^2/L$
near $ \rho_c $ for several system sizes and $ r=0.5$,
in Figs.~4(a), (b), and (c), respectively.
We see the clear trend that the transition becomes sharper
with increasing the system size.
Such size effects at a glance resemble
to that near the phase transition point
of equilibrium critical phenomena.

\bigskip
\leftline{3.2. {\large \sl Thermodynamic Limit}}\label{sec:thermo}
\medskip

In this subsection, we discuss the model
in the thermodynamic limit $ L \to \infty $ and
the long-time limit $T \to \infty$.

First, we deal with the pure rule-184 model, that is, $ r=1.0 $.
Let $ \rho_p$ be the particle concentration and
$\rho_h=1-\rho_p$ be the hole concentration.
For $ \rho_p > \rho_h $,
we expect that the movable particle number per one step
coincides asymptotically with the hole number:
\eqn{ \lim_{T\to\infty} \frac{1}{T} \sum_i
=\rho_h L . }
Then
\eqn{
v=\frac{\rho_h L}{n}=\frac{\rho_h}{\rho_p}=\frac{1-\rho_p}{\rho_p}.
\label{eq:ansjam}
}
For $ \rho_p < \rho_h $, on the other hand,
we expect that the movable particle number is just the particle
number:
\eqn{
\lim_{T\to\infty}\frac{1}{T} \sum_i n_i =\rho_p .}
Then
\eqn{
v=\frac{\rho_p L}{n}=1.
\label{eq:ansfree}
}
Consequently, for the rule-184 CA
we get
\eqn{
v  =&1 & \quad \mbox{($ 0<\rho<1/2$)} \non \\
  = & \disp \frac{1-\rho}{\rho} & \quad \mbox{($ 1/2 < \rho <1$) .}
\label{eq:anspeed}
}
It is well known that the above result is exact not only for
the inifinite system
but also for any finite system of the rule-184 CA.

Next we consider the blockage model.
In the flow-constant phase,
the two regions, that is, the free region and the fully-jam region
coexist as have been seen in the previous section.
According to the observations in the previous section,
we assume that these two regions are locally equivalent to the
corresponding phases in pure rule-184 CA;
Especially the particle concentration is assumed uniform
in both regions.
Let $ \rho_f$ and $\rho_j$ be the concentration in the free
region and the concentration in the jam region, respectively.
For the particle number to conserve, the relation
\eqn{
 \rho L=\rho_j h +\rho_f (L-h) \label{eq:cpn}
}
should hold in the flow-constant phase.
According to the discussion we have already made on the
``life time'' of the particle on the blocakge site,
the particle concentration in the free region is
\eqn{
\rho_f=\frac{r}{1+r}. \label{eq:cf}
}
Because of the particle-hole symmetry of the model,
the transmission of the particle into the free region is equivalent
to the transmission of the hole into the jam region.
Thus the above discussion also apply to the jam region,
and the hole concentration there coincides with $\rho_f$;
The particle-hole symmetry in the jam region is equivalent to
\eqn{
\rho_f+\rho_j=1. \label{eq:cf2}
}
{}From eqs.~(\ref{eq:cf}) and ~(\ref{eq:cf2}), we determine $\rho_f$
and $\rho_j$ as
\eqn{
\rho_f=\frac{r}{1+r}, \qquad  \disp \rho_j = &\disp \frac{1}{1+r}.
\label{eq:den}
}
Therefore both the concentration in the free region and
the jam region are independent of the total density $\rho$;
Rather, they depend only on the transmission rate $r$.

{}From eq.~(\ref{eq:cpn}), we obtain the width of the jam region as
\eqn{
\frac{h}{L}= \frac{\rho-\rho_f}{\rho_j-\rho_f} .
\label{eq:width2}
}
Combining eq.~(\ref{eq:den}) and eq.~(\ref{eq:width2}), we get
\eqn{
\frac{h}{L} = \frac{(1+r) \rho-r}{1-r}.
\label{eq:width3}
}
The critical concentration $\rho_c$ is determined from
eq.~(\ref{eq:width3})
by putting $h=0$ and $h=L$;
We get $ \rho_c = \rho_f $ and  $ \rho_j$.
Thus by solving eq.~(\ref{eq:den}),
the phase boundary in the $(\rho, r)$-plane in the thermodynamic
limit is obtained as
\eqn{
r=\frac{\rho}{1-\rho}, \qquad \mbox{\rm and} \qquad  r= \disp
\frac{1-\rho}{\rho}.
\label{eq:phased}
}
The phase diagram thus obtained is shown in Fig.~5.


The average speed in the flow-constant phase can also be calculated.
Since the dynamics of the particles is locally rule-184,
the particles in the free region move with the speed $ v_f=1 $,
and the particles in the jam region move with the speed
$ v_j= (1-\rho_j)/\rho_j$.
Thus the average speed of the particles in the flow-constant phase
can be written as
\eqn{
v=\frac{L}{ \disp
\frac{\De}{v_j}+\frac{L-\De}{v_f}}, \label{eq:speed2}
}
which gives
\eqn{
v=\frac{r}{1+r} \frac{1}{\rho}.
\label{eq:speed3}
}

Putting above considerations together,
we get the average speed in the thermodynamic limit:
\eqn{
v  = \left\{
\begin{array}{@{\,} ll}
1 & \qquad \mbox{($ 0 <\rho < \rho_f$) }  \\
\disp \frac{r}{1+r} \frac{1}{\rho} &  \qquad \mbox{($
\rho_f < \rho < \rho_j $) }  \\
\disp \frac{1-\rho}{\rho} &  \qquad \mbox{($ \rho_j <
\rho < 1 $),}
\end{array}
 \right.
\label{eq:sumspeed}
}
with $\rho_f=r/(1+r)$ and $\rho_j = 1/(1+r)$.
We also get the width of the jam region,
\eqn{
h/L= \left\{
\begin{array}{@{\,} ll}
0 &  \qquad \mbox{($ 0 <\rho < \rho_f$ )}  \\
\disp \frac{\rho-\rho_f}{\rho_j-\rho_f} & \qquad
\mbox{($ \rho_f < \rho < \rho_j $ )}
\\
1 & \qquad  \mbox{($ \rho_j < \rho < 1 $).}
\end{array}
\right.
\label{eq:sumwidth}
}

Since the local motion of the particles in the both regions are
treated exactly by applying the results of the rule-184 CA,
we expect that the above results are exact in the thermodynamic
limit. For finite systems, on the other hand,
the effects of randomness at the bottleneck
will cause deviations in local motion from the pure rule-184 CA near
the bottleneck;
Such deviations, however, is expected to remain only in some small
region near the bottleneck,
so that they do not affect the properties in the thermodynamic limit.
In fact, the solid lines shown in Fig.~1(a), which represent the
relation eq.~(\ref{eq:sumspeed}),
very well to the simulation data,
apart from the region near the critical concentration;
Especially, the flow-constant phase is very well described by
eq.~(\ref{eq:sumspeed}).

\bigskip
\leftline{3.3. {\large \sl Finite-Size Scaling}}\label{sec:scaling}
\medskip

In Figs. ~4(a)--(c),
clear size effects are observed near the critical concentration.
As was discussed in the last subsection,
such finite-size effects are the consequences of the randomness at
the bottleneck. Our next step is to study the nature of these
finite-size effects.
To this end, we carry out analyses based on
the finite-size scaling method,
which is widely used in the field of the critical phenomena.

First, we consider the finite-size scaling of the average speed.
Suppose $v$ vanishes with the system size $L$ at the critical point
with a power law:
\eqn{
 1-v \sim L^\xi,
\label{eq:sbe}
}
where we introduced a scaling dimension $\xi$ for the average speed.
Following the usual procedure for the finite-size scaling analysis,
we assume $1-v$ is expressed in a scaling form
\eqn{
1-v=L^\xi f_v \left( \frac{\rho-\rho_c}{
\rho_c L^\zeta } \right),
\label{eq:scspeed}
}
where $f_v$ is an unknown scaling function whose argument
is the reduced concentration $(\rho -\rho_c)/\rho_c$ scaled by
the system size.
The exponent $\zeta$ is the scaling exponent for the reduced
concentration.
For the concentration much larger than $\rho_c$
(in the off-critical region),
the flow is expected to approach the constant value $\rho_c$.
If we take that into account,
we have to assign the same exponent to both the average speed and
the reduced concentration, that is, we put $\xi=\zeta$.
Thus the expected scaling relation for $v$ becomes
\eqn{
1-v & = & L^\xi f_v \left( \frac{\rho-\rho_c}{ \rho_c L^\xi }
\right).
\label{eq:scsp1}
}

Next we consider the finite-size scaling for the width of the jam
region.
The normalized width $h/L$ is also expected to vanish
in the thermodynamic limit at $\rho_c$;
Thus we suppose again the power-law dependence of $h/L$ on $L$:
\eqn{
\frac{h}{L} \sim L^{\phi},
\label{eq:widthpower}
}
where $\phi$ is the scaling exponent for the normalized width.
Let us assume the scaling form similar to eq.~(\ref{eq:scsp1}):
\eqn{
\frac{h}{L}=L^\phi f_h \left( \frac{\rho-\rho_c}{\rho_c L^\xi}
\right),
\label{eq:sch}
}
where $f_h$ is an unknown scaling function.
We expect $h$ is order $L$ at $ (\rho-\rho_c)/\rho_c \gg 1$,
and thus expect
\eqn{
\frac{h}{L}=const.
\label{eq:scaling2}
}
To take that into account, we put $ \phi=\xi $.
Therefore $v$ and $h/L$ scale with the same exponent.
The expected scaling relation for $h/L$ thus becomes
\eqn{
\frac{h}{L}=L^\xi f_h \left( \frac{\rho-\rho_c}{\rho_c L^\xi}
\right).
\label{eq:scaling3}
}

Finally, from a simple dimension counting, we expect
that the scaling relation for the fluctuation of the width is written
as
\eqn{
\frac{\De h^2}{L} \sim L^{2 \xi+1}f_{\De h}
\left(\frac{\rho-\rho_c}{\rho_c L^\xi} \right)
\label{eq:fluctbe}
}
with an unknown scaling function $f_{\De h}$.
It should be noted that only a single scaling exponent $\xi$
appears in these scaling relations.
If the scaling width $h/L$ at $\rho_c$ is simply a consequence of
random fluctuations,
it may behaves as $L^{-1/2}$;
Therefore, from eq.~(\ref{eq:widthpower}) we expect $\xi = -1/2$
for random fluctuation.

We investigate the validity of the scaling relations derived above
with the simulation data.
We show results only for the transmission rate $r=0.5$ in the
following.
Figure ~6 shows the scaled average speed
$(1-v ) L^{1/2} $ against the scaled reduced concentration.
We used $ (\rho-\rho_c)/\rho $ as the reduced concentration
instead of $ (\rho-\rho_c)/\rho_c $ used in eq.~(\ref{eq:scsp1}),
just because we can get better scaling plot with the former than
with the latter.
Such change of the definition for the reduced concentration, however,
does not affect the leading scaling behavior;
it, in fact, only gives an analytic correction-to-scaling.
We see in the scaling plot that all the data for different system
sizes indeed collapsed onto a single scaling function.
For $ \rho-\rho_c >0 $,
this scaling function seems to approach a linear function of
$(\rho-\rho_c) L^{1/2} /\rho $.
This behavior is consistent with the fact that the flow takes the
constant
value $ \rho_c $.

In Fig.~7, we plot the scaled width of
jam phase $hL^{1/2}$ against the scaled reduced concentration.
We took $ (\rho-\rho_c) L^{1/2} $ for the horizontal axis.
Again we see that all the data fall on a single curve.
The scaling function seems to approach the linear
function of $ (\rho-\rho_c) L^{1/2} $, for $ \rho-\rho_ c > 0 $;
That again is consistent to the fact
that $h/L$ in the flow-constant phase is proportional to $
\rho-\rho_c $.

So far, we have seen that the simulation data are consistent
with $\xi = -1/2$.
Putting this value into eq.~(\ref{eq:fluctbe}),
the scaling relation for the fluctuation of width becomes
\eqn{
\frac{\De h^2}{L} \sim L^{0}f_{\De h} \left(\frac{\rho-\rho_c}{\rho_c
L^{-1/2}} \right).
\label{eq:fluctbe2}
}
In cases of usual critical phenomena, the power $L^0$ implies
the logarithmic divergence.
In the present case, however, where $\xi=-1/2$ is a consequence of
the random fluctuation,
the power $L^0$ may imply simply that $\De h^2/L$ does not depend on
$L$.
In Fig.~8, we plot the fluctuation of width $\De h^2/L$
without rescale by $L$
against the scaled reduced concentration.
We see a somewhat different scaling behavior
from those of $v$ or $h/L$:
While all the data for the concentration lower than the peak
collapse into a single curve,
the data for higher concentration do not scale.
Thus the scaling relation eq.~(\ref{eq:fluctbe2}) holds
only for the concentration lower than the peak,
which appears at a small positive value of the reduced concentration.
The size dependence of $\De h^2/L$ for higher concentration
can be deduced from the raw plot in Fig.~4(c).
We can see that $\De h^2 /L $ for the concentration higher than the
peak is almost independent of the system size
without rescaling of the concentration.
This behavior in this region coincides with the result
for TASEP with blockage,
where the unnormalized fluctuation of the shock-front position
in the mixed phase behaves as $L^{1/2}$,\cite{JL92}
although the definition of the interface position is
different between two models.

\section{Summary and Discussion}\label{sec:sum}
\setcounter{equation}{0}

We have examined a one-dimensional cellular automaton model of
the traffic flow with a bottleneck,
by introducing a blockage into the rule-184 CA.
We found the formation of the jam region after the bottleneck
when the car concentration exceeds some critical value determined
by the transmission rate of the bottleneck.
As a result, in the intermedeate range of the concentration between
free phase and the fully-jam phase,
that is, the flow-constant phase,
the coexistence of free region and the fully-jam region is observed
with well-defined boundary between them.
Such behaviors agree with our experience in real traffic ways,
when road constructions or tunnels exist.
In fact, we sometimes find ourselves trapped in a traffic jam all of
a sudden,
after having driven freely.
In the fully-jam region after the blockage of the present model,
a car does not move constantly;
Rather it moves and stops alternatively.
Such intermittent motion also agree with our real experience.
Thus the model treated in the present study can represent some
realities despite its simple structure.

We have discussed the properties of the traffic flow
in the thermodynamic limit,
where the fluctuations due to the randomness at the blockage
can be ignored.
The expressions we obtained for the average velocity and the average
flow
reproduce the simulation results very accurately
apart from the concentration near the critical value.
Since the local motions of the particles are treated exactly by
applying the known results of the rule-184 CA,
the above expressions are expected to be exact
in the thermodynamicl limit.
We also obtained the phase diagram of the present model
in terms of the concentration and the transmission rate.

By the computer simulation of the model,
the strong finite-size effects near the transition point was
observed,
which were analysed by means of the finite-size scaling.
Since the pure rule-184 model does not exhibit such finite-size
effects,
they are induced by a randomness only at the single blockage
introduced in the system.
Although the finite-size behaviors of the present model
resemble to those of the equilibrium critical phenomena,
they are caused simply by the random fluctuations.
It should be noted that no such finite-size behaviors have been
reported
so far for the stochastic two-lane model,\cite{N94}
although the overall behavior is close to the present model.

Exact solutions have been found for some TASEP-related models.
\cite{derrida1,derrida2}
After having completed the present work,
we came to know that quite recently
Janowsky and Lebowitz obtained the exact solutions on small lattices
for the same model as they treated in ref.~13,
which is a stochastic TASEP model with a blockage,
and extrapolated them to the thermodynamic limit by
Pad\'e approximation.\cite{JL93}
As another related model in the context of TASEP,
Sch\"utz have found the generalized-Bethe-ansatz solution for the
two-sublattice deterministic version of the TASEP with a
blockage,\cite{schutz}
where the system is divided into two sublattices and
the particles in the same sublattice move simultaneously.
Although these three models including the present CA-based model
resemble with each other at a glance,
their update dynamics are quite different.
As a consequence, they behave differently in detail;
Especially the phase boundaries do not coincide with each other.
More detailed comparison between the models will be studied in near
future.

\acknowledgement{
We are grateful to Y.~Akutsu for valuable discussions.
We also thank S.~A. Janowsky for sending us ref.~16
prior to the publication.
The work was partially supported by a Grant-in-Aid
for Scientific Research on Priority Areas, ``Computational Physics
as a New Frontier in Condensed Matter Research'', from the Ministry
of Education, Science and Culture, Japan.}

\newpage

%
%
%
\newpage
{\large Figure Captions}
\begin{description}

\item[Fig. ~1(a) ]
Typical results for average speed, 99 sites; transmission
rate r=0.3, 0.5, 1.0. Solid lines are solutions of
mean-field-like analysis.
\item[Fig. ~1(b)]
Typical results for flow, 99 sites; transmission rate
r= 0.3, 0.5, 1.0.
\item[Fig. ~2(a) ]
Typical results for width of the jam region, 99 sites;
transmission rate r=0.3,0.5.
\item[Fig. ~2(b)]
Typical results for fluctuations of width of the jam
region, 99 sites; transmission rate r=0.3, 0.5.
\item[Fig. ~3(a)]
Snap shot for 100 site and transmission rate r=0.5 at
density 0.5. White square is particle and blockage site is 3.
\item[Fig. ~3(b)]
Snap shot for 100 site and transmission rate r=0.5 at
density 0.33, near the critical concentration. White
square is particle and blockage site is 3.
\item[Fig. ~4(a)]
Average speed for several system size; transmission
rate r=0.5; near the critical density $ \rho_c =1/3$.
\item[Fig. ~4(b)]
Width of a jam phase for several system size;
transmission rate r=0.5; near the critical density
$ \rho_c =1/3$.
\item[Fig. ~4(c)]
Fluctuations of the width for several system size;
transmission rate r=0.5; near the critical density
$ \rho_c =1/3$.
\item[Fig. ~5]
Phase diagram for infinite system with blockage;
Density $\rho$ vs transmission rate $ r$.
\item[Fig. ~6]
Scaled average speed for several system size;
transmission rate r=0.5; near the critical density.
$ \rho_c=1/3$.
\item[Fig. ~7]
Scaled width of jam phase for several system size;
transmission rate is 0.5; near the critical density
$ \rho_c=1/3$.
\item[Fig. ~8]
Scaled fluctuations for several system size;
transmission rate is 0.5; near the critical density
$ \rho_c=1/3$.
\end{description}
\end{document}